\documentclass[12pt,oneside]{article}

\begin{document}
\thispagestyle{empty}
\setcounter{page}{0}
\renewcommand{\theequation}{\thesection.\arabic{equation}}

\vspace{2cm}

\begin{center}
{\bf Emergent Spacetime}

\vspace{1.4cm}

Nathan Seiberg

\vspace{.2cm}

{\em School of Natural Sciences, } \\
{\em Institute for Advanced Study,} \\
{\em Princeton, NJ 08540 USA} \\
\end{center}

\vspace{-.1cm}

\centerline{{\tt seiberg@ias.edu}} \vspace{1cm}
\centerline{ABSTRACT}

\vspace{- 4 mm}  

\begin{quote}\small
We summarize the arguments that space and time are likely to be
emergent notions; i.e.\ they are not present in the fundamental
formulation of the theory, but appear as approximate macroscopic
concepts. Along the way we briefly review certain topics. These
include ambiguities in the geometry and the topology of space
which arise from dualities, questions associated with locality,
various known examples of emergent space, and the puzzles and the
prospects of emergent time.
\end{quote}
\baselineskip18pt
\noindent

\vspace{5cm}

\begin{quote}\small Rapporteur talk at the 23rd Solvay Conference
in Physics, December, 2005.
\end{quote}

\newpage

\setcounter{equation}{0}
\section{Introduction}

The purpose of this talk is to review the case for the idea that
space and time will end up being emergent concepts; i.e.\ they
will not be present in the fundamental formulation of the theory
and will appear as approximate semiclassical notions in the
macroscopic world. This point of view is widely held in the string
community and many of the points which we will stress are well
known.

Before we motivate the idea that spacetime should be emergent, we
should discuss the nature of space in string theory.  We do that
in section 2, where we review some of the ambiguities in the
underlying geometry and topology. These follow from the dualities
of string theory. T-duality leads to ambiguities at the string
length $l_s$ and the quantum dualities lead to ambiguities at the
Planck length $l_p \ll l_s$. All these ambiguities in the geometry
are associated with the fact that as we try to probe the space
with increasing resolution, the probes we use become big and
prevent us from achieving the desired accuracy.

The discussion about ambiguities in space will lead us to make
some comments about locality. In particular, we will ask whether
to expect locality in a space or in one of its duals.

In section 3 we will briefly mention some of the peculiar
non-gravitational theories which are found as certain limits of
string theory.  Some of them are expected to be standard field
theories, albeit without a Lagrangian.  Others, like theories on a
noncommutative space or little string theory, are not local
quantum field theory.  They exhibit interesting nonlocal behavior.

In section 4 we will make the case that general covariance is
likely to be a derived concept.

Section 5 will present several examples of emergent space. First
we will discuss the simplest examples which do not involve
gravity. Then we will turn to four classes of examples of emergent
space: the emergent two-dimensional (worldsheet) gravity from the
matrix model, the celebrated gauge/gravity duality, linear dilaton
backgrounds, and the BFSS matrix model.  We will discuss some of
their properties and will stress the similarities and the
differences between them. In particular, we will discuss their
finite temperature behavior as a diagnostic of the system in
extreme conditions.

Section 6 will be devoted to emergent time.  Here we do not have
concrete examples.  Instead, we will present some of the
challenges and confusions that this idea poses.  We will also
mention that understanding how time emerges will undoubtedly shed
new light on some of the most important questions in theoretical
physics including the origin of the Universe.

We will summarize the talk in section 7 where we will also present
some general speculations.

Before we start we should mention some important disclaimers.  As
we said, most of the points which will be discussed here are
elementary and are well known in the string community.  We
apologize for boring you with them. Other points will be
inconclusive because they reflect our confusions. Also, not all
issues and all points of view will be presented. Instead, the
presentation will be biased by my prejudice and my own work.  For
example, the discussion will focus on string theory (for
textbooks, see \cite{gsw, Polchinski}), and other approaches to
quantum gravity will not be reviewed.  Since this talk is expected
to lead to a discussion, we will present certain provocative and
perhaps outrageous ideas. Finally, there will be very few
references, mostly to reviews of the subject, rather than to
original papers.


\setcounter{equation}{0}
\section{Ambiguous space}

\subsection{Ambiguous space in classical string theory}

We start this section by discussing the ambiguities in the
geometry and the topology which exist already at string tree
level. These are usually referred to as {\it T-duality} (for
reviews, see e.g.\ \cite{T-duality, Sen:2001di}).

Consider strings propagating in some background fields (e.g.\
metric).  Clearly, these background fields should satisfy the
equations of motion.  Then, it turns out that different
backgrounds can lead to the same physics without any observable
difference between them.  Therefore, there is no unique answer to
the question: {\it ``What is the background metric?"} and the
background geometry is ambiguous.

Intuitively, these ambiguities arise from the extended nature of
the string.  Features in the geometry which are smaller than the
string length $l_s = \sqrt{\alpha'}$ cannot be detected using a
string probe whose characteristic size is $l_s$.\footnote{D-branes
\cite{Polchinski} which are smaller than $l_s$ can sometime lead
to a more precise metric, but different kinds of D-branes lead to
different answers and therefore the ambiguity is not resolved.}

The simplest and most widely known example of this ambiguity is
the equivalence between a circle with radius $R$ and a circle with
radius $\alpha'/R$.  A slightly more peculiar example is the
equivalence between a circle with radius $R=2\sqrt{\alpha'}$ and a
$\mathcal{Z}_2$ quotient of a circle (a line segment) with
$R=\sqrt{\alpha'}$.  This example demonstrates that even the
topology is ambiguous.  Furthermore, we can start with a circle of
radius $R$, smoothly change it to $R=2\sqrt{\alpha'}$, then use
the duality with the line segment and then change the length of
the line segment.  This way we start with a circle which is not
dual to a line segment and we continuously change its topology to
a line segment which is not dual to a circle.

A characteristic feature of these dualities is the role played by
momentum and winding symmetries.  In the example of the two
circles with radii $R$ and $\alpha'/R$ momentum conservation in
one system is mapped to winding conservation in the other.
Momentum conservation arises from a geometric symmetry (an
isometry) of the circle.  It is mapped to winding conservation
which is a {\it stringy symmetry.}  This is a manifestation of the
stringy nature of T-duality and it makes it clear that it is
associated with the extended nature of the string.

In some situations there exists a description of the system in
terms of a {\it macroscopic background}; i.e.\ the space and all
its features are larger than $l_s$.  This is the most natural
description among all possible dual descriptions. However, two
points should be stressed about this case. First, even though this
description is the most natural one, there is nothing wrong with
all other T-dual descriptions and they are equally valid. Second,
it is never the case that there is more than one such macroscopic
and natural description.

More elaborate and richer examples of this fundamental phenomenon
arise in the study of Calabi-Yau spaces.  Here two different
Calabi-Yau spaces which are a ``mirror pair'' (for a review, see
e.g.\ \cite{mirror}) lead to the same physics.  Furthermore, it is
often the case that one can continuously interpolate between
different Calabi-Yau spaces with different topology.  These
developments had dramatic impact on mathematics (see e.g.
\cite{mirror, Dijkgraaf}).

Another kind of T-duality is the cigar/Sine-Liouville duality
\cite{FZZ}.  One side of the duality involves the cigar geometry:
a semi-infinite cylinder which is capped at one side. It has a
varying dilaton, such that the string coupling at the open end of
the cigar vanishes.  This description makes it clear that the
shift symmetry around the cigar leads to conserved momentum.
However, the string winding number is not conserved, because wound
strings can slip through the capped end of the cigar.  The other
side of this duality involves an infinite cylinder.  Here the
winding conservation is broken by a condensate of wound strings.
The cigar geometry is described by a two-dimensional field theory
with a nontrivial metric but no potential, while its dual, the
Sine-Liouville theory, is a theory with a flat metric but a
nontrivial potential. This example again highlights the importance
of the winding modes. It also demonstrates that the T-duality
ambiguity is not limited to compact dimensions.  Here the
ambiguity is between two different non-compact systems (an
infinite and a half infinite cylinder).

From the worldsheet point of view T-duality represents an exact
equivalence between different two-dimensional conformal field
theories.  Therefore, the phenomenon of T-duality persists beyond
classical string theory, and extends to all orders in perturbation
theory.  Furthermore, in some situations one can argue that
T-duality is a gauge symmetry.  This observation means that
T-duality is exact and it cannot be violated non-perturbatively.

The phenomenon of T-duality leads us to ask two interesting
questions.  First, is $l_s$ a minimum length; i.e.\ is the notion
of distance ill defined below $l_s$?  Second, is the theory local
in one space, or in its T-dual space, or in neither?  We will
return to these questions below.

Before we leave the topic of ambiguities in classical string
theory we would like to mention another important stringy
phenomenon which is associated with the extended nature of the
string.  The high energy density of string states is such that the
canonical ensemble of free strings does not exist above a certain
temperature $T_H\sim {1\over l_s}$, which is known as the {\it
Hagedorn temperature} \cite{gsw, Polchinski}.  The relevant modes
which lead to this phenomenon are long strings. They have large
entropy and hence the partition function diverges at $T_H$.
Equivalently, when Euclidean time is compactified on a circle of
radius $R={1\over 2 \pi T}$ (with thermal boundary conditions) an
instability appears when $R\le {1\over 2 \pi T_H}$. This
instability is associated with strings which are wound around the
Euclidean time circle. $T_H$ could be a limiting temperature,
beyond which the theory does not exist. Alternatively, this
phenomenon could mean that the system undergoes a first order
phase transition to another phase. That phase could exhibit the
fundamental degrees of freedom more clearly.  Again we see that
the theory tries to hide its short distance behavior.

\subsection{Ambiguous space in quantum string theory}

Quantum mechanics introduces new ambiguities in space which are
related to new dualities (for reviews, see e.g.\ \cite{Polchinski,
Sen:2001di}).  These ambiguities go beyond the obvious ambiguities
due to the quantum fluctuations. Here the characteristic length
scale is the Planck length $l_p \ll l_s$.

An intuitive argument explaining the origin of these ambiguities
is the following.  If we want to explore space with resolution of
order $r$, the uncertainly principle tells us that we need to use
energy $E> {1\over r} $. This energy has to be concentrated in a
region of size $r$.  But in the presence of gravitational
interactions, this concentration of energy creates a black hole
unless $r>l_p$. Therefore, {\it we cannot explore distances
smaller than the Planck length}.

It is important to stress that although the ambiguities in the
quantum theory are often described as of different nature than the
ambiguities in the classical theory, fundamentally they are quite
similar.  Both of them are associated with the breakdown of the
standard small distance/high energy connection -- as we try to
increase the energy of a probe it becomes bigger and does not
allow us to explore short distances.

The {\it quantum dualities}, which are also known as S-duality or
U-duality, extend the classical T-duality and lead to a beautiful
and coherent picture of stringy dualities.  These exchange highly
quantum situations with semiclassical backgrounds, exchange
different branes, etc.  As in the classical dualities, among all
dual descriptions there is at most one description which is
natural because it is semiclassical.  All other dual descriptions
are very quantum mechanical.

\subsection{Comments about locality}

We now turn to some comments about locality in string theory.

Quantum field theory is local.  This locality guarantees that the
theory is causal.  We would like string theory also to be causal
or at least macroscopically causal.  Furthermore, we know that at
long distances string theory behaves like quantum field theory and
therefore it is macroscopically local.  But is string theory local
also over short distances?

One piece of evidence in favor of locality is the analyticity of
the perturbative string S-matrix.  Normally, causality and
locality lead to analyticity.  Since the string S-matrix is
analytic, it is likely that string theory is local. However, it is
logically possible that a slightly weaker condition than locality
and therefore of causality can also guarantee the analyticity of
the S-matrix.

One reason string theory might not be local in a standard way is
the extended nature of the interacting objects, the strings.  At
the most naive and intuitive level locality of string interactions
is not obvious.  Even though two strings interact at a point to
form a third string, this interaction is nonlocal when viewed from
the point of view of the center of masses of the interacting
strings. It is known that this nonlocality is harmless and is
consistent with the analyticity of the S-matrix.\footnote{In open
string field theory a basis based on the string midpoint replaces
the basis based on the center of mass and then the interaction
appears to be local.}

We would like to comment about locality and the cosmological
constant. The old fashioned point of view of the cosmological
constant problem suggested that its value is related to some kind
of a UV/IR mixing and to violation of naive locality -- the short
distance theory somehow reacts to long distance fluctuations and
thus sets the value of the cosmological constant.  A more modern
point of view on the subject is that the cosmological constant is
set anthropically (see, e.g.\ \cite{Polchinskitalk}).  It remains
to be seen whether the cosmological constant is a hint about some
intrinsic nonlocality in the theory.

The ambiguities we discussed above might hint at some form of
nonlocality.  We have stressed that increasing the energy of a
probe does not lead to increased resolution.  Instead, the probe
becomes bigger and the resolution is reduced.  This point is at
the heart of the various dualities and ambiguities in the
background.  We have already asked whether we expect locality in a
space, or in its dual space. It is hard to imagine that the theory
can be simultaneously local in both of them. Then, perhaps it is
local in neither.  Of course, when a macroscopic weakly coupled
natural description exists, we expect the theory to be at least
approximately local in that description.

It is important to stress that although intuitively the notion of
locality is obvious, this is not the case in string theory or in
any generally covariant theory.  The theory has no local
observables.  Most of the observables are related to the S-matrix
or other objects at infinity.  These do not probe the detailed
structure of the theory in the interior.  Therefore, without local
observables it is not clear how to precisely define locality.

We will argue below that space and time should be emergent
concepts. So if they are not fundamental, the concept of locality
cannot be fundamental as well.  It is possible that locality will
end up being ill defined, and there will be only an approximate
notion of locality when there is an approximate notion of
spacetime.

\setcounter{equation}{0}
\section{Non-standard theories without gravity}

Next, let us digress slightly to review some of the non-standard
theories without gravity that were found by studying various
limits of string theory.  These theories exhibit interesting and
surprising new phenomena.  We expect that these theories and their
peculiar phenomena will be clues to the structure of the
underlying string theory.  Since they are significantly simpler
than string theory, they could be used as efficient laboratories
or toy models.

The first kind of surprising theories are new local field theories
which cannot be given a standard Lagrangian description. These are
superconformal field theories in five or six dimensions with
various amount of supersymmetry.  The most symmetric examples are
the six-dimensional (2,0) theories (for a review, see e.g.\
\cite{Witten:2002eh}).  They are found by taking an appropriate
scaling limit of string theory in various singularities or on
coincident 5-branes.  The existence of these theories calls for a
new formulation of local quantum field theory without basing it on
a Lagrangian.

Another class of interesting non-gravitational theories are {\it
field theories on noncommutative spaces} (for a review, see e.g.\
\cite{Douglas:2001ba}). These theories do not satisfy the standard
rules of local quantum field theory.  For example, they exhibit a
UV/IR mixing which is similar to the UV/IR mixing in string theory
-- as the energy of an object is increased its size becomes
bigger.

The most enigmatic theories which are derived from string theory
are the {\it little string theories} (for a review, see e.g.\
\cite{Aharony:1999ks}).  These non-gravitational theories exhibit
puzzling stringy behavior.  The stringy nature of these theories
arises from the fact that they appear by taking a certain scaling
limit of string theory (in the presence of NS5-branes or some
singularities) while keeping $\alpha'$ fixed. One stringy
phenomenon they exhibit is T-duality. This suggests that despite
the lack of gravity, these theories do not have a local energy
momentum tensor. Otherwise, there should have been several
different energy momentum tensors which are related by T-duality.
It was also argued that because of their high energy behavior
these theories cannot have local observables. Finally, these
theories exhibit Hagedorn spectrum with a Hagedorn temperature
which is below $T_H$ of the underlying string theory. It was
suggested that this Hagedorn temperature is a limiting
temperature; i.e.\ the canonical ensemble does not exist beyond
that temperature.

\setcounter{equation}{0}
\section{Derived general covariance}

The purpose of this section is to argue that general covariance
which is the starting point of General Relativity might not be
fundamental.  It could emerge as a useful concept at long
distances without being present in the underlying formulation of
the theory.

General covariance is a {\it gauge symmetry}.  As with other gauge
symmetries, the term ``symmetry'' is a misnomer.  Gauge symmetries
are not symmetries of the Hilbert space; the Hilbert space is
invariant under the entire gauge group.  Instead, gauge symmetries
represent a redundancy in our description of the theory.  (It is
important to stress, though, that this is an extremely useful
redundancy which allows us to describe the theory in simple local
and Lorentz invariant terms.)

Indeed, experience from duality in field theory shows that gauge
symmetries are not fundamental.  It is often the case that a
theory with a gauge symmetry is dual to a theory with a different
gauge symmetry, or no gauge symmetry at all. A very simple example
is Maxwell theory in 2+1 dimensions.  This theory has a $U(1)$
gauge symmetry, and it has a dual description in terms of a free
massless scalar without a local gauge symmetry.  More subtle
examples in higher dimensions were found in supersymmetric
theories (for reviews, see e.g.\ \cite{Seiberg:1995ac,
Intriligator:1995au}).

If ordinary gauge symmetries are not fundamental, it is reasonable
that general covariance is also not fundamental.  This suggests
that the basic formulation of the theory will not have general
covariance.  General covariance will appear as a derived (and
useful) concept at long distances.

An important constraint on the emergence of gauge symmetries
follows from the Weinberg-Witten theorem \cite{Weinberg:1980kq}.
It states that if the theory has massless spin one or spin two
particles, these particles are gauge particles.  Therefore, the
currents that they couple to are not observable operators.  If
these gauge symmetries are not present in some formulation of the
theory, these currents should not exist there.   In particular, it
means that if an ordinary gauge symmetry emerges, the fundamental
theory should not have this symmetry as a global symmetry.  In the
context of emergent general covariance, this means that the
fundamental theory cannot have an energy momentum tensor.

If we are looking for a fundamental theory without general
covariance, it is likely that this theory should not have an
underlying spacetime. This point is further motivated by the fact
that General Relativity has no local observables and perhaps no
local gauge invariant degrees of freedom. Therefore, there is
really no need for an underlying spacetime.  Spacetime and general
covariance should appear as approximate concepts which are valid
only macroscopically.

\setcounter{equation}{0}
\section{Examples of emergent space}

\subsection{Emergent space without gravity}

The simplest examples of emergent space are those which do not
involve gravity.  Here the starting point is a theory without a
fundamental space, but the resulting answers look approximately
like a theory on some space. The first examples of this kind were
the {\it Eguchi-Kawai} model and its various variants (for a
review, see e.g.\ \cite{Makeenko:2002uj}).  Here a $d$ dimensional
$SU(N)$ gauge theory is formulated at one point. The {\it large
$N$} answers look like a gauge theory on a macroscopic space.

Certain extensions of the (twisted) Eguchi-Kawai model are
theories on a {\it noncommutative space} (for a review, see e.g.\
\cite{Douglas:2001ba}). Here the coordinates of the space do not
commute and are well defined only when they are macroscopic.

A physical realization of these ideas is the {\it Myers effect}
\cite{Myers:1999ps}.  Here we start with a collection of $N$
branes in some background flux.  These branes expand and become a
single brane of higher dimension.  The new dimensions of this
brane are not standard dimensions.  They form a so-called ``fuzzy
space.''  In the {\it large $N$ limit} the resulting space becomes
macroscopic and its fuzzyness disappears.

\subsection{Emergent space with gravity: matrix model of 2d
gravity}

The first examples of emergent space with gravity and general
covariance arose from the {\it matrix model of random surfaces}
(for a review, see e.g.\ \cite{Ginsparg:1993is}).  Here we start
with a certain matrix integral or matrix quantum mechanics and
study it in perturbation theory.  Large Feynman diagrams of this
perturbation expansion can be viewed as discretized
two-dimensional surfaces.

This system is particularly interesting when the size of the
matrices $N$ is taken to infinity together with a certain limit of
the parameters of the matrix integral.  In this double scaling
limit the two-dimensional surfaces become large and smooth and the
system has an effective description in terms of random surfaces.
The degrees of freedom on these surfaces are local quantum fields
including a dynamical metric and therefore this description is
generally covariant.

The formulation of these theories as matrix models does not have a
two-dimensional space nor does it have general covariance. These
concepts emerge in the effective description.

In addition to being interesting and calculable models of
two-dimensional gravity, these are concrete examples of how space
and its general covariance can be emergent concepts.

\subsection{Emergent space with gravity: Gauge/Gravity duality}

The most widely studied examples of emergent space with gravity
are based on the AdS/CFT correspondence \cite{Maldacena:1997re,
Gubser:1998bc, Witten:1998qj, Aharony:1999ti}. This celebrated
correspondence is the duality between string theory in AdS space
and a conformal field theory at its boundary. Since other speakers
in this conference will also talk about it, we will only review it
briefly and will make a few general comments about it.

The bulk theory is a theory of gravity and as such it does not
have an energy momentum tensor.  The dual field theory on the
boundary has an energy momentum tensor.  This is consistent with
the discussion above about emergent gravity (section 4), because
the energy momentum tensor of the field theory is in lower
dimensions than the bulk theory and reflects only its boundary
behavior.

The operators of the boundary theory are mapped to string states
in the bulk.  A particularly important example is the energy
momentum tensor of the boundary theory which is mapped to the bulk
graviton. The correlation functions of the conformal field theory
are related through the correspondence to string amplitudes in the
AdS space. (Because of the asymptotic structure of AdS, these are
not S-matrix elements.) When the field theory is deformed by
relevant operators, the background geometry is slightly deformed
near the boundary but the deformation in the interior becomes
large.   This way massive field theories are mapped to nearly AdS
spaces.

The radial direction in AdS emerges without being a space
dimension in the field theory.  It can be interpreted as the
renormalization group scale, or the energy scale used to probe the
theory.  The asymptotic region corresponds to the UV region of the
field theory.  This is where the theory is formulated, and this is
where the operators are defined.  The interior of the space
corresponds to the IR region of the field theory.  It is
determined from the definition of the theory in the UV.

A crucial fact which underlies the correspondence, is the infinite
warp factor at the boundary of the AdS space.  Because of this
warp factor, finite distances in the field theory correspond to
infinite distances in the bulk. Therefore, a field theory
correlation function of finitely separated operators is mapped to
a gravity problem which infinitely separated sources.

An important consequence of this infinite warp factor is the
effect of finite temperature.  The boundary field theory can be
put at finite temperature $T$ by compactifying its Euclidean time
direction on a finite circle of radius $R={1 \over 2 \pi T}$.  At
low temperature, the only change in the dual asymptotically AdS
background it to compactify its Euclidean time. Because of the
infinite warp factor, the radius of the Euclidean time circle in
the AdS space is large near the boundary, and it is small only in
a region of the size of the AdS radius $R_{AdS}$. Therefore, most
of the bulk of the space is cold.  Only a finite region in the
interior is hot.  As the system is heated up, the boundary theory
undergoes a thermal deconfinement phase transition.  In the bulk
it is mapped to the appearance of a Schwarzschild horizon at small
radius and the topology is such that the Euclidean time circle
becomes contractible.  For a CFT on a 3-sphere, this phase
transition is the Hawking-Page transition, and the dual high
temperature background is AdS-Schwarzschild.  Both above and below
the transition the bulk asymptotes to (nearly) AdS.  Most of it
remains cold and it is not sensitive to the short distance
behavior of string theory.

While the boundary field theory is manifestly local, locality in
the bulk is subtle.  Because of the infinite warp factor, possible
violation of locality in the bulk over distances of order $l_s$
could be consistent with locality at the boundary.  In fact, it is
quite difficult to find operators in the field theory which
represent events in the bulk which are localized on scales of
order $R_{AdS}$ or smaller.  This underscores the fact that it is
not clear what we mean by locality, if all we can measure are
observables at infinity.

These developments have led to many new insights about the two
sides of the duality and the relation between them (for a review,
see \cite{Aharony:1999ti}).  In particular, many new results about
gauge theories, including their strong coupling phenomena like
thermal phase transitions, confinement and chiral symmetry
breaking were elucidated. The main new insight about gravity is
its {\it holographic nature} -- the boundary theory contains all
the information about the bulk gravity theory which is higher
dimensional.  Therefore, the number of degrees of freedom of a
gravity theory is not extensive.  This is consistent with the lack
of local observables in gravity.

\subsection{Emergent space with gravity: linear dilaton
backgrounds}

\subsubsection{Generalities}

Another class of examples of an emergent space dimension involves
backgrounds with a linear dilaton direction.  The string coupling
constant depends on the position in the emergent direction,
parameterized by the spatial coordinate $\phi$, through $g_s(\phi)
= e^{Q\phi \over 2}$ with an appropriate constant $Q$.  Therefore,
the string coupling constant vanishes at the boundary $\phi \to
-\infty$. The other end of the space at $\phi \to +\infty$ is
effectively compact.

Like the AdS examples, here the bulk string theory is also dual to
a theory without gravity at the boundary.   In that sense, this is
another example of holography.  However, there are a few important
differences between this duality and the AdS/CFT duality.

In most of the linear dilaton examples the holographic theory is
not a standard local quantum field theory.  For example, the near
horizon geometry of a stack of NS5-branes is a linear dilaton
background which is holographic to the little string theory (for a
review, see e.g.\ \cite{Aharony:1999ks}).  The stringy, non-field
theoretic nature of the holographic theory follows from the fact
that it has nonzero $\alpha'$, and therefore it exhibits
T-duality.

Because of the vanishing interactions at the boundary of the
space, the interactions take place in an effectively compact
region (the strong coupling end).  Therefore, we can study the
S-matrix elements of the bulk theory.  These are the observables
of the boundary theory.

Unlike the AdS examples, the string metric does not have an
infinite warp factor.  Here finite distances in the boundary
theory correspond to finite distances (in string units) in the
bulk. Therefore, it is difficult to define {\it local} observables
in the boundary theory and as a result, the holographic theory is
not a local quantum field theory.

This lack of the infinite warp factor affects also the finite
temperature behavior of the system.  Finite temperature in the
boundary theory is dual to finite temperature in the entire bulk.
Hence, the holographic theory can exhibit Hagedorn behavior and
have maximal temperature.

\subsubsection{Matrix model duals of linear dilaton backgrounds}

Even though the generic linear dilaton theory is dual to a
complicated boundary theory, there are a few simple cases where
the holographic theories are very simple and are given by the
large $N$ limit of certain matrix models.

The simplest cases involve strings in one dimension $\phi$ with a
linear dilaton. The string worldsheet theory includes a Liouville
field $\phi$ and a $c<1$ minimal model (or in the type 0 theory a
$\hat c <1$ superminimal model).  The holographic description of
these {\it minimal string theories} is in terms of the large $N$
limit of matrix integrals (for a review, see e.g.\
\cite{Seiberg:2004at}).

Richer theories involve strings in two dimensions: a linear
dilaton direction $\phi$ and time $x$ (for a review, see e.g.\
\cite{Klebanov:1991qa}). Here the holographic theory is the large
$N$ limit of matrix quantum mechanics.

These two-dimensional string theories have a finite number of
particle species.  The bosonic string and the supersymmetric 0A
theory have one massless boson, and the 0B theory has two massless
bosons. Therefore, these theories do not have the familiar
Hagedorn density of states of higher dimensional string theories,
and correspondingly, their finite temperature behavior is smooth.

One can view the finite temperature system as a system with
compact Euclidean time $x$.  Then, the system has $R \to
\alpha'/R$ T-duality which relates high and low temperature.  As a
check, the smooth answers for the thermodynamical quantities
respect this T-duality.

It is important to distinguish the two different ways matrix
models lead to emergent space.  Above (section 5.2) we discussed
the emergence of the two-dimensional string worldsheet with its
worldsheet general covariance.  Here, we discuss the target space
of this string theory with the emergent holographic dimension
$\phi$.

Since the emergence of the holographic direction in these systems
is very explicit, we can use them to address various questions
about this direction. In particular, it seems that there are a
number of inequivalent ways to describe this dimension.  The most
obvious description is in terms of the Liouville field $\phi$. A
second possibility is to use a free worldsheet field which is
related to $\phi$ through a nonlocal transformation (similar to
T-duality transformation). This is the Backlund field of Liouville
theory. A third possibility, which is also related to $\phi$ in a
nonlocal way arises more naturally out of the matrices as their
eigenvalue direction. These different descriptions of the emergent
direction demonstrate again that the ambiguity in the description
of space which we reviewed above (section 2) is not limited to
compact dimensions. It also highlights the question of locality in
the space.  In which of these descriptions do we expect the theory
to be local? Do we expect locality in one of them, or in all of
them, or perhaps in none of them?

\subsubsection{2d heterotic strings}

We would like to end this subsection with a short discussion of
the heterotic two-dimensional linear dilaton system.  Even though
there is no known holographic matrix model dual of this system,
some of its peculiar properties can be analyzed.

As with the two-dimensional linear dilaton bosonic and type 0
theories, this theory also has a finite number of massless
particles.  But here the thermodynamics is more subtle.  We again
compactify Euclidean time on a circle of radius $R$.  The
worldsheet analysis shows that the system has $R \to \alpha'/2R$
T-duality. Indeed, the string amplitudes respect this symmetry.
However, unlike the simpler bosonic system, here the answers are
not smooth at the selfdual point $R=\sqrt{\alpha'/2}$.  This lack
of smoothness is related to long macroscopic strings excitations
\cite{Seiberg:2005nk}.

What is puzzling about these results is that they cannot be
interpreted as standard thermodynamics.  If we try to interpret
the Euclidean time circle as a thermal ensemble with temperature
$T={1\over 2\pi R}$, then the transition at $R=\sqrt{\alpha'/2}$
has negative latent heat.  This violates standard thermodynamical
inequalities which follow from the fact that the partition
function can be written as a trace over a Hilbert space ${\rm Tr}\
e^{-H/T}$ for some Hamiltonian $H$.  Therefore, we seem to have a
contradiction between compactified Euclidean time and finite
temperature.  The familiar relation between them follows from the
existence of a Hamiltonian which generates {\it local time
evolution}. Perhaps this contradiction means that we cannot
simultaneously have locality in the circle and in its T-dual
circle.  For large $R$ the Euclidean circle answers agree with the
thermal answers with low temperature.  But while these large $R$
answers can be extended to smaller $R$, the finite temperature
interpretation ceases to make sense at the selfdual point.
Instead, for smaller $R$ we can use the T-dual circle, which is
large, and describe the T-dual system as having low temperature.

\subsection{Emergent space in the BFSS matrix model}

As a final example of emergent space we consider the BFSS matrix
model (for a review, see e.g.\ \cite{Banks:1999az}).  Its starting
point is a large collection of D0-branes in the lightcone frame.
The lightcone coordinate $x^+$ is fundamental and the theory is an
ordinary quantum mechanical system with $x^+$ being the time.

The transverse coordinates of the branes $x^i$ are the variables
in the quantum mechanical system.  They are not numbers.  They are
$N$ dimensional matrices.  The standard interpretation as
positions of the branes arises only when the branes are far apart.
Then the matrices are approximately diagonal and their eigenvalues
are the positions of the branes.  In that sense the transverse
dimensions emerge from the simple quantum mechanical system.

The remaining spacetime direction, $x^-$, emerges holographically.
It is related to the size of the matrices $N \sim p_-$ where $p_-$
is the momentum conjugate to $x^-$.

\setcounter{equation}{0}
\section{Emergent time}

After motivating the emergence of space it is natural to ask
whether time can also emerge.  One reason to expect it is that
this will put space and time on equal footing -- if space emerges,
so should time.  This suggests that time is also not fundamental.
The theory will be formulated without reference to time and an
approximate (classical) notion of macroscopic time, which is our
familiar ``time'', will emerge.  Microscopically, the notion of
time will be ill defined and time will be fuzzy.

There are several obvious arguments that time should not be
emergent:
\begin{enumerate}
    \item Even though we have several examples of emergent space,
    we do not have a single example of emergent time.
    \item We have mentioned some of the issues associated with
    locality in emergent space.  If time is also emergent we are
    in danger of violating locality in time and that might lead to
    violation of causality.
    \item It is particularly confusing what it means to have a theory
    without fundamental time.  Physics is about predicting the
    outcome of an experiment {\it before} the experiment is performed.
    How can this happen without fundamental time and without
    notions of ``before and after''?  Equivalently, physics is
    about describing the evolution of a system.  How can systems
    evolve without an underlying time?   Perhaps these
    questions can be avoided, if some order of events
    is well defined without an underlying time.
    \item More technically, we can ask how much of the standard
    setup of quantum mechanics should be preserved.  In particular,
    is there a wave function?  What is its probabilistic
    interpretation?  Is there a Hilbert space of all possible wave
    functions, or is the wave function unique?  What do we mean by
    unitarity (we cannot have unitary evolution, because without
    time there is no evolution)?  Some of these questions are
    discussed in \cite{Hartle}.
\end{enumerate}

My personal prejudice is that these objections and questions are
not obstacles to emergent time.  Instead, they should be viewed as
challenges and perhaps even clues to the answers.

Such an understanding of time (or lack thereof) will have, among
other things, immediate implications for the physics of space-like
and null singularities (for a review, see e.g.\ \cite{Gibbons})
like the black hole singularity and the cosmological singularity.
We can speculate that understanding how time emerges and what one
means by a wave function will explain the meaning of {\it the
wave-function of the Universe.}  Understanding this wave function,
or equivalently understanding the proper initial conditions for
the Universe, might help resolving some of the perplexing
questions of vacuum selection in string theory.  For a review of
some aspects of these questions see \cite{Polchinskitalk}.

\setcounter{equation}{0}
\section{Conclusions and speculations}

We have argued that spacetime is likely to be an emergent concept.
The fundamental formulation of the theory will not have spacetime
and it will emerge as an approximate, classical concept which is
valid only macroscopically.

One challenge is to have emergent spacetime, while preserving some
locality --  at least macroscopic locality, causality,
analyticity, etc. Particularly challenging are the obstacles to
formulating physics without time.  It is clear that in order to
resolve them many of our standard ideas about physics will have to
be revolutionized. This will undoubtedly shed new light on the
fundamental structure of the theory.

Understanding how time emerges will also have other implications.
It will address deep issues like the cosmological singularity and
the origin of the Universe.

We would like to end this talk with two general speculative
comments.

Examining the known examples of a complete formulation of string
theory, like the various matrix models, AdS/CFT, etc., a
disturbing fact becomes clear. It seems that many different
definitions lead to a consistent string theory in some background.
In particular, perhaps every local quantum field theory can be
used as a boundary theory to define string theory in (nearly) AdS
space. Perhaps every quantum mechanical system can be the
holographic description of string theory in 1+1 dimensions. And
perhaps even every ordinary integral defines string theory in one
Euclidean dimension.  With so many different definitions we are
tempted to conclude that we should not ask the question:\ {\it
``What is string theory?''} Instead, we should ask:\ {\it ``Which
string theories have macroscopic dimensions?''} Although we do not
have an answer to this question, it seems that {\it large $N$}
will play an important role in the answer.

Our second general comment is about reductionism -- the idea that
science at one length scale is derived (at least in principle)
from science at shorter scales.  This idea has always been a theme
in all branches of science. However, if there is a basic length
scale, below which the notion of space (and time) does not make
sense, we cannot derive the principles there from deeper
principles at shorter distances. Therefore, once we understand how
spacetime emerges, we could still look for more basic fundamental
laws, but these laws will not operate at shorter distances.  This
follows from the simple fact that  the notion of ``shorter
distances'' will no longer make sense.  This might mean the {\it
end of standard reductionism.}

\vspace{5mm}

\noindent {\bf Acknowledgments}: We would like to thank the
organizers of the 23rd Solvay Conference in Physics for arranging
such an interesting and stimulating meeting and for inviting me to
give this talk. We also thank T.~Banks, I.~Klebanov, J.~Maldacena,
and D.~Shih for useful comments and suggestions about this
rapporteur talk. This research is supported in part by DOE grant
DE-FG02-90ER40542.

\vspace{5mm}

%

\end{document}